\documentclass[conference]{IEEEtran}
\IEEEoverridecommandlockouts
\usepackage{cite}
\usepackage{amsmath,amssymb,amsfonts}
\usepackage{algorithmic}
\usepackage{graphicx}
\usepackage{textcomp}
\usepackage{xcolor}
\def\BibTeX{{\rm B\kern-.05em{\sc i\kern-.025em b}\kern-.08em
    T\kern-.1667em\lower.7ex\hbox{E}\kern-.125emX}}

\usepackage{./note_pdf}

\begin{document}

\bstctlcite{IEEEexample:BSTcontrol}

\title{Parametric Modeling and Estimation of Photon Registrations for 3D Imaging\\
\thanks{The work is supported, in part, by the DARPA / SRC CogniSense JUMP 2.0 Center, NSF IIS-2133032, and NSF ECCS-2030570.}
}

\author{\IEEEauthorblockN{Weijian Zhang, Hashan K. Weerasooriya, Prateek Chennuri, and Stanley H. Chan}
\IEEEauthorblockA{\textit{School of Electrical and Computer Engineering} \\
\textit{Purdue University}\\
West Lafayette, USA \\
Emails: \{zhan5056, hweeraso, pchennur, stanchan\}@purdue.edu}
}

\maketitle

\begin{abstract}
In single-photon light detection and ranging (SP-LiDAR) systems, the histogram distortion due to hardware dead time fundamentally limits the precision of depth estimation. To compensate for the dead time effects, the photon registration distribution is typically modeled based on the Markov chain self-excitation process. However, this is a discrete process and it is computationally expensive, thus hindering potential neural network applications and fast simulations. In this paper, we overcome the modeling challenge by proposing a continuous parametric model. We introduce a Gaussian-uniform mixture model (GUMM) and periodic padding to address high noise floors and noise slopes respectively. By deriving and implementing a customized expectation maximization (EM) algorithm, we achieve accurate histogram matching in scenarios that were deemed difficult in the literature.
\end{abstract}


\section{Introduction}
Single-photon avalanche diode (SPAD) is one of the emerging single-photon sensitive sensors~\cite{rochas2003single, charbon_spad-based_2013, fossum_quanta_2016, ma_review_2022} with tens-of-picosecond time resolution. Time-correlated single-photon counting (TCSPC) is an advanced technique for rapidly measuring photon responses~\cite{beckerAdvancedTimeCorrelatedSingle2005}. An acquisition system combining these two tools greatly outperforms traditional CMOS or CCD cameras in various imaging applications such as fluorescence lifetime imaging (FLIM), non-line-of-sight (NLOS) imaging, and single-photon light detection and ranging (SP-LiDAR)~\cite{bruschini_single-photon_2019, faccio_non-line--sight_2020, massa_laser_1997, rapp_advances_2020}.

The combination of SPAD and TCSPC particularly benefits SP-LiDAR in precisely identifying objects from long distances and in low-light conditions~\cite{Li:20}. By actively illuminating the scene with a periodic train of narrow laser pulses of a known shape, an SP-LiDAR system time-tags the delays of reflecting photons and gradually builds a histogram throughout all repetition periods $t_r$ for each pixel. It is well-known that photon arrivals can be modeled by a Poisson process~\cite{snyderRandomPointProcesses1991}. The histogram of arriving photons is proportional to the photon arrival flux function, a shifted version of the transmitted pulse, and can be used for depth estimation through a matched filter.

However, not every arriving photon can be recorded due to the hardware limit. Independent dead times exist in SPAD and TCSPC during which a SPAD cannot detect and a TCSPC module cannot register a photon respectively. The resulting photon loss distorts the empirical histogram, impeding the accurate depth estimation. Therefore, it is critical to describe the photon registration process for further distortion corrections.

Modeling the photon registration process has been a subject of interest for more than a decade, but existing methods are either only tailored for classic LiDAR systems or based on the Markov chain self-excitation process~\cite{rappDeadTimeCompensation2019, Rapp:21}, which is computationally intense and discrete. Our goal in this paper is to \textit{characterize the photon registration probability density function (PDF) in modern LiDAR systems by using a simple and continuous parametric model}. We use the expectation maximization (EM) algorithm~\cite{baumMaximizationTechniqueOccurring1970} to estimate the model parameters. The predicted PDF will enable us to run a matched filter and estimate the depths~\cite{rappDeadTimeCompensation2019}.

\subsection{Sources of Distortions}
Different types of SPADs and TCSPC mechanisms lead to different distortions. In this work, we assume the distortions caused by nonparalyzable SPADs and modern TCSPC modules, the same setup as in~\cite{rappDeadTimeCompensation2019}.

The dead time of a nonparalyzable SPAD is a fixed constant $t_d$. Whereas, for a paralyzable SPAD, it will extend for at least another $t_d$ if a new photon hits the sensor during the current dead time\cite{fellerRETRACTEDCHAPTERProbability2015}. Although an independent dead time $t_e$ exists in TCSPC, we assume $t_d > t_e$ so that we can ignore $t_e$ according to~\cite{isbanerDeadtimeCorrectionFluorescence2016}.

Based on the synchronization mechanism, we can classify typical TCSPC schemes into classic and modern TCSPC~\cite{beckerAdvancedTimeCorrelatedSingle2005}. A comparison is depicted in Fig.~\ref{fig: TCSPC_Schemes}.

\subsubsection{Classic TCSPC}
The laser sync starts the timer and the first photon detection stops it. The time difference is recorded as the time delay, and the dead time follows. Chances are that the end of the dead time is in the middle of a repetition period and a new photon arrival is detected by the SPAD in the current period. However, the TCSPC cannot register detections until the next repetition period, which is labeled in purple in Fig.~\ref{fig: TCSPC_Schemes}. Therefore, the acquisition system can register at most one photon per cycle. The empirical histogram is expected to shift to the left because the system is prone to record earlier photons, known as ``classic pile-up"~\cite{beckerAdvancedTimeCorrelatedSingle2005}.

\subsubsection{Modern TCSPC}
Different from classic TCSPC, a modern TCSPC module has independent timers for the laser sync and the detection signal, bringing the potential of multiple registrations in one period\cite{wahlDeadtimeOptimizedTimecorrelated2007}. The disconnection between the deactivation and the laser sync makes the distortions more complex. We call them ``dead time distortions" in this paper.

\begin{figure}[tbp]
\centering
\includegraphics[width=1\linewidth]{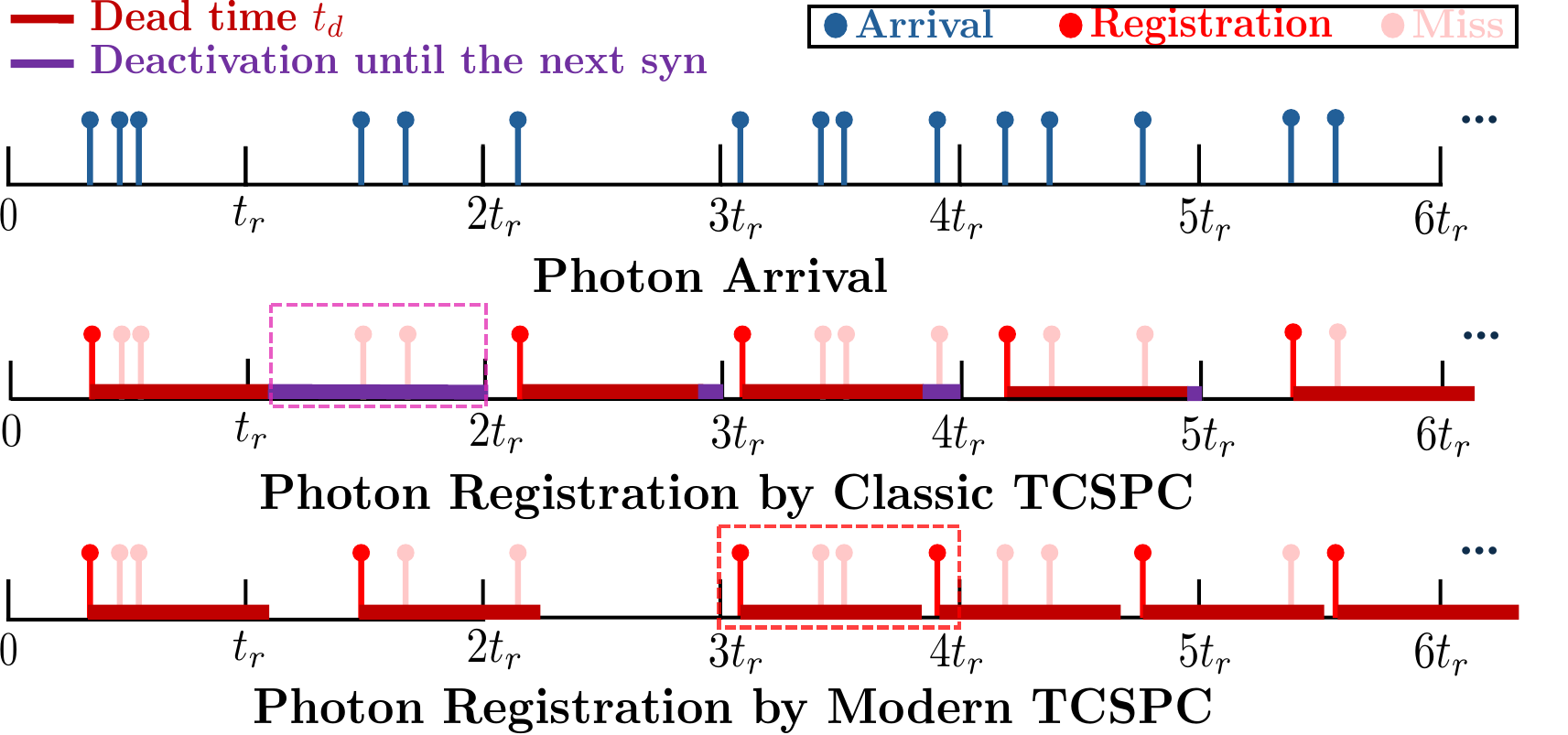}
\caption{Comparison between classic and modern TCSPC.}
\label{fig: TCSPC_Schemes}
\end{figure}

From Fig.~\ref{fig: TCSPC_Schemes}, the reactivation of registration in classic TCSPC begins along with the laser sync each time, enforcing at most one registration per cycle. The registered photons are thus independent. By contrast, the reactivation in modern TCSPC is right after the end of the dead time, which is hard to predict. This scheme shortens the deactivation time and speeds up the acquisition rate ($2$ registrations within $(3t_r, 4t_r]$). However, the uncertainty complicates the distortion shapes. As pile-up effects have been studied since~\cite{coatesCorrectionPhotonPileup1968}, we consider the more difficult dead time distortions in this work.

\subsection{Related Work}

Existing modeling and estimation methods can be divided into low-flux regimes and high-flux regimes where people aim at pile-up and dead time distortions separately.

\subsubsection{Low-flux Regime}
A typical and universal way to deal with the distortion is to circumvent it by constraining the photon level such that the detection frequency is at most 5\%  of all repetition cycles~\cite{yu_mean_2000,phillips_time_1985}. Under this flux condition, the photon level is so low that the probability that a photon reaches the sensor during the dead time is almost $0$. Thus, the empirical histogram will match the photon flux function. However, the time required to collect enough photons for further tasks such as depth estimation becomes longer, which is not ideal for some applications demanding fast acquisition rates. Given this, a series of works are devoted to reducing the minimum number of photons needed for precise estimation~\cite{kirmani_first-photon_2014, shin_photon-efficient_2015, rapp_few_2017, altmann_lidar_2016}. Nevertheless, our method is operated in the high-flux regime with a fundamentally faster acquisition rate.

\subsubsection{Pile-up}
Moving from the low-flux regime, a number of teams have recently been working on deriving signal-processing-based algorithms \cite{pediredlaSignalProcessingBased2018, heideSubpicosecondPhotonefficient3D2018, guptaPhotonFloodedSinglePhoton3D2019, guptaAsynchronousSinglePhoton3D2019, poAdaptiveGatingSinglePhoton2022} to compensate for distortions in the high-flux regime. However, only classic LiDAR systems and thus pile-up effects have been examined in the aforementioned literature. In this case, the registered photons are statistically independent and can be modeled as a Poisson multinomial distribution~\cite{daskalakisStructureCoveringLearning2015}. By contrast, modeling the dead time distortions is much more challenging, but it will enable faster acquisition than the classic setup.

\subsubsection{Dead Time Distortions}
\begin{figure}[tbp]
\centering
\includegraphics[width=1\linewidth]{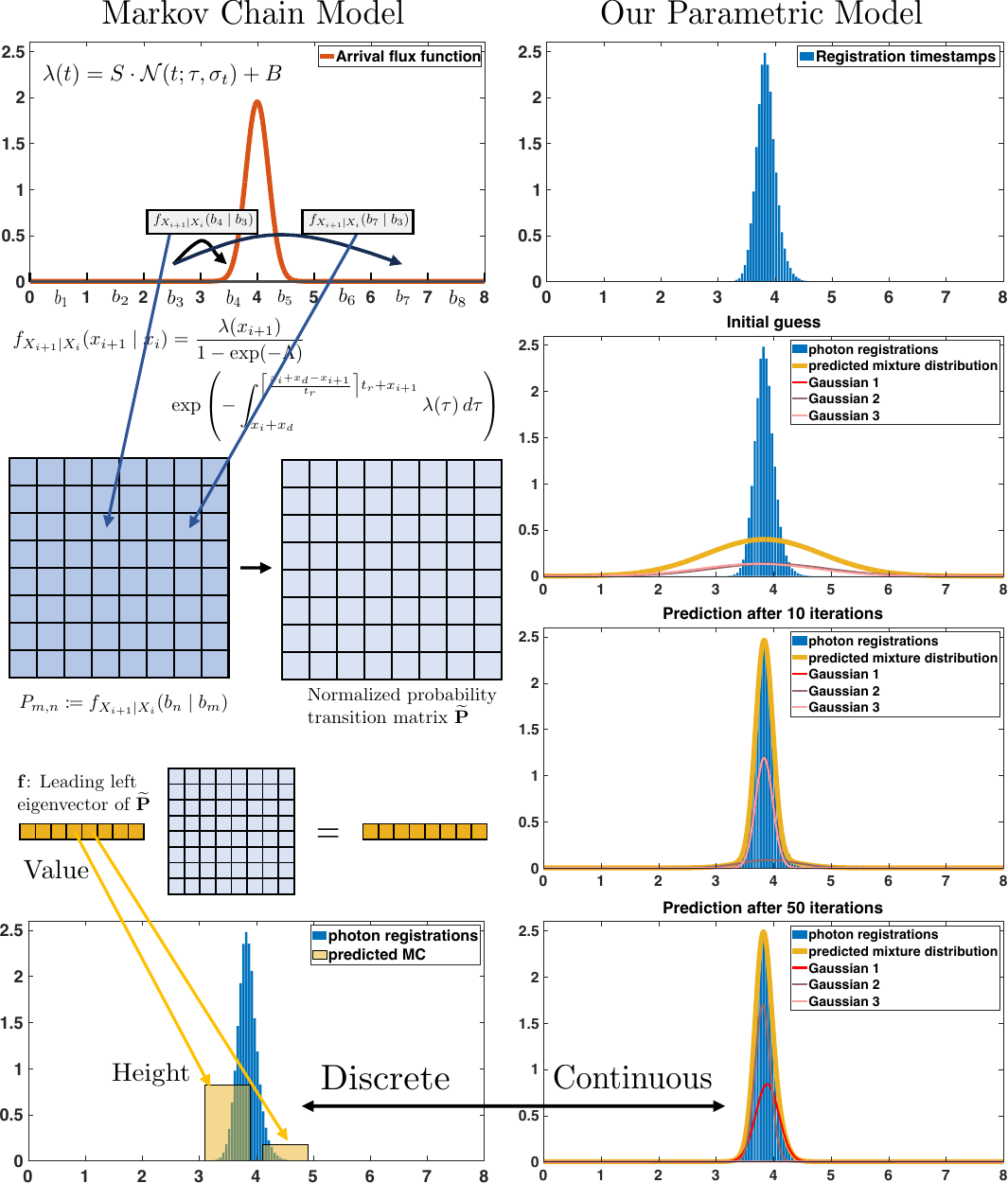}
\caption{Markov chain model vs. our parametric model.}
\label{fig: MC comparison}
\end{figure}

In modern LiDAR systems, consecutive photon registrations become dependent. In~\cite{isbanerDeadtimeCorrectionFluorescence2016}, the photon registration process is identified as a time-evolving attenuation of the photon arrival flux function, and the attenuation is computed iteratively. One step further, \cite{rappDeadTimeCompensation2019, Rapp:21} rigorously model the photon registration process as a Markov chain. The PDF is approximated by the limiting Markov chain stationary PDF, obtained by calculating the leading left eigenvector $\vf$ of the discretized probability transition matrix $\widetilde{\mP}$. Although the reported prediction is accurate, it is discrete and computationally heavy. A comparison between the Markov chain model and our parametric model is illustrated in Fig.~\ref{fig: MC comparison}. We notice an intrinsic trade-off between the binning resolution, thus prediction accuracy, and the computation complexity.

\subsection{Main Contributions}
In this paper, we develop a unified framework for the scenarios above. Our contributions are summarized as follows.
\subsubsection{}
We introduce a parametric model for the timestamp distribution in the existence of dead time. Based on a mixture of Gaussian and uniform density, our model can accurately characterize the distribution with at most eighteen parameters. Compared to the discrete Markov chain model~\cite{rappDeadTimeCompensation2019}, this is a substantial reduction, thus allowing us to simulate photon registration timestamps much faster. For inverse problems, our continuous model is both differentiable for deep learning and optimizable for maximum likelihood (ML) estimation.

\subsubsection{}
We introduce the EM algorithm to estimate the model parameters. We introduce a periodic padding idea to handle the intricate distortion shapes such as the noise slope. We show that the estimated parameters can accurately reproduce the histogram of the timestamps.

\section{Background}
In this section, we provide the background of the photon arrival process in SP-LiDAR and present a method to simulate the distortions caused by dead time.

\subsection{Photon Arrival Process}
Our photon arrival model follows \cite{Chan_2024_CVPR}. For each repetition cycle at each pixel $(i, j)$, the photon arrival is characterized by an inhomogeneous Poisson process with the flux intensity function
\begin{equation}
    \lambda_{i,j}(t) = \alpha_{i,j} \cdot s(t - \tau) + \lambda_b,
\end{equation}
where $\alpha_{i,j}$ is the object reflectivity at this pixel, $\lambda_b$ is the constant background noise across all pixels, $\tau$ is the pulse delay implying the true depth, and $s$ describes the shape of the transmitted pulse. We choose the shape to be Gaussian, i.e. $s(t - \tau) = \calN(t; \tau, \sigma_t)$ where $\sigma_t$ denotes the half width of the Gaussian pulse.

For simplicity, we drop the indexing $(i,j)$ and make the following assumptions. The transmitted pulse is bounced back from only one location. The repetition cycle $t_r$ is wide enough to contain $\tau$ inside. The Gaussian pulse width is very narrow, i.e. $\sigma_t \ll t_r$. Thus, $\int_0^{t_r} s(t - \tau) \, dt = 1$. We model the sensor quantum efficiency constant $\eta$ and the dark current constant $\lambda_d$ as parts of $\alpha$ and $\lambda_b$, respectively.

Within one cycle, the energy carried by $\lambda(t)$ is
\begin{equation}
    Q \bydef \int_0^{t_r} \lambda(t) \, dt = \alpha + t_r\lambda_b = S + B,
\end{equation}
where we define the signal level $S := \alpha$ and the noise level $B := t_r\lambda_b$. $Q$ indicates the mean number of photon arrivals each cycle.

\subsection{Simulation Scheme}
To simulate the photon arrival process, we follow the simulation strategy of photon arrival timestamps from \cite{Chan_2024_CVPR}, which allows us to quickly collect a data cell $\mX_A = [ \vx_1, \vx_2, \ldots, \vx_K ]$ after $K$ repetition cycles. Each $\vx_k$ $(k=1, 2, \ldots, K)$ stores a timestamp vector for that cycle, with possibly different lengths or even empty. Data in $\mX_A$ are called relative photon arrival timestamps since they are all within $[0, t_r)$.

\begin{figure}[tbp]
\centering
\includegraphics[width=1\linewidth]{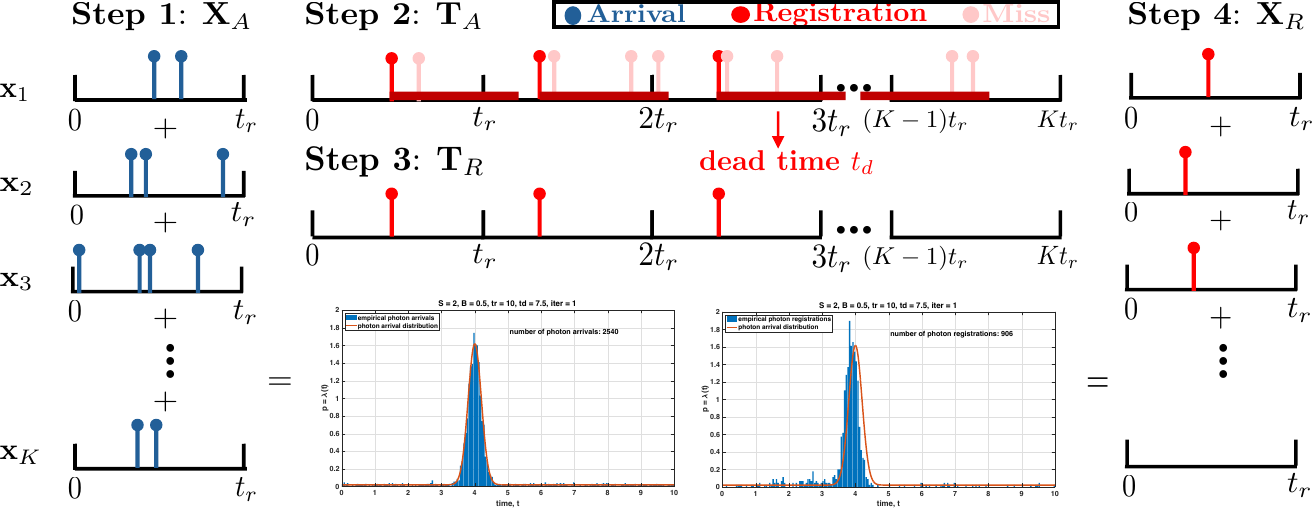}
\caption{Illustration of dead time distortion simulation.}
\label{fig: Dead time simulation illustration}
\end{figure}

However, in the presence of dead time, some arriving photons will not be recorded. To determine whether a photon should be eliminated, we order the timestamps in $\mX_A$ sequentially across $K$ cycles. Each value $X_A$ in $\mX_A$ is transformed into an absolute photon arrival timestamp $T_A$ according to $T_A = (k-1) \cdot t_r + X_A$. Then, we compare the time differences between consecutive photon arrivals with $t_d$ and keep the photons outside dead time as absolute photon registration timestamps $\mT_R$. Finally, we transform each value $T_R$ in $\mT_R$ back to the relative photon registration using $X_R = \texttt{mod}(T_R, t_r)$. All $X_R$'s can be binned to build a histogram. The procedures of simulating one photon registration histogram are summarized in Fig.~\ref{fig: Dead time simulation illustration}. Due to the photon loss, the histogram is distorted toward the left.

\begin{figure}[tbp]
     \centering
     \subfloat[]{\includegraphics[width=0.24\textwidth]{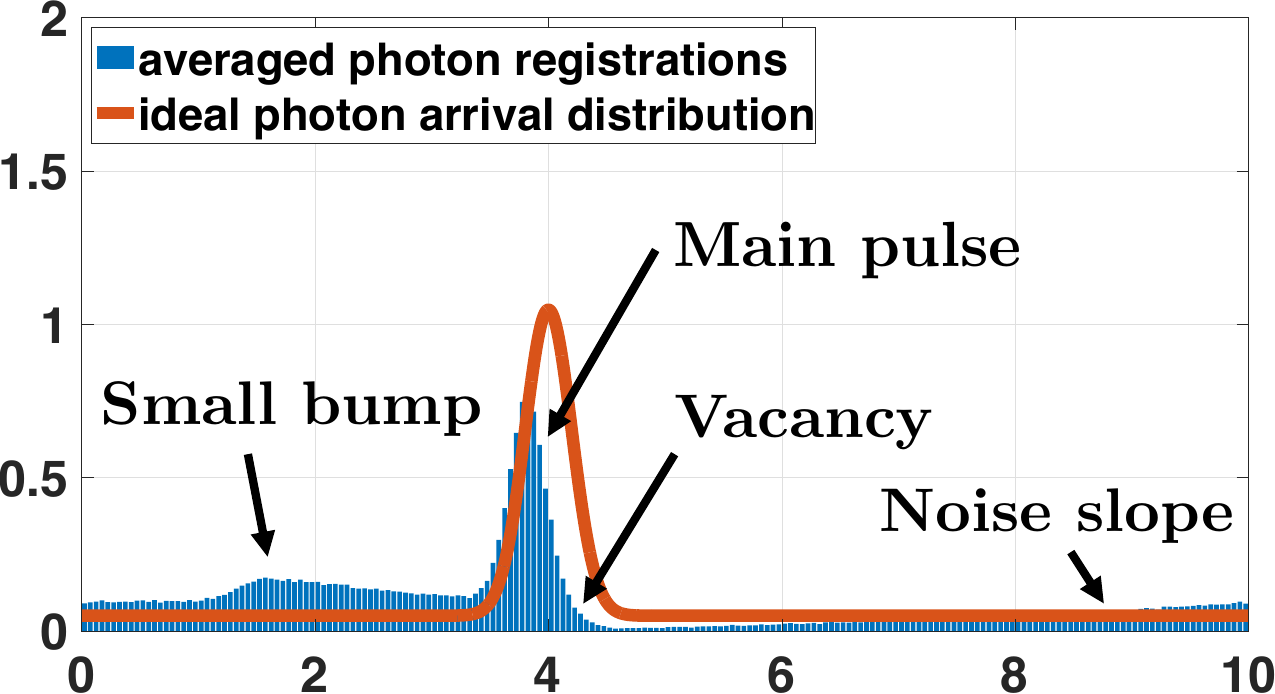}\label{fig: dead time histogram (a)}}
     \hfill
     \subfloat[]{\includegraphics[width=0.24\textwidth]{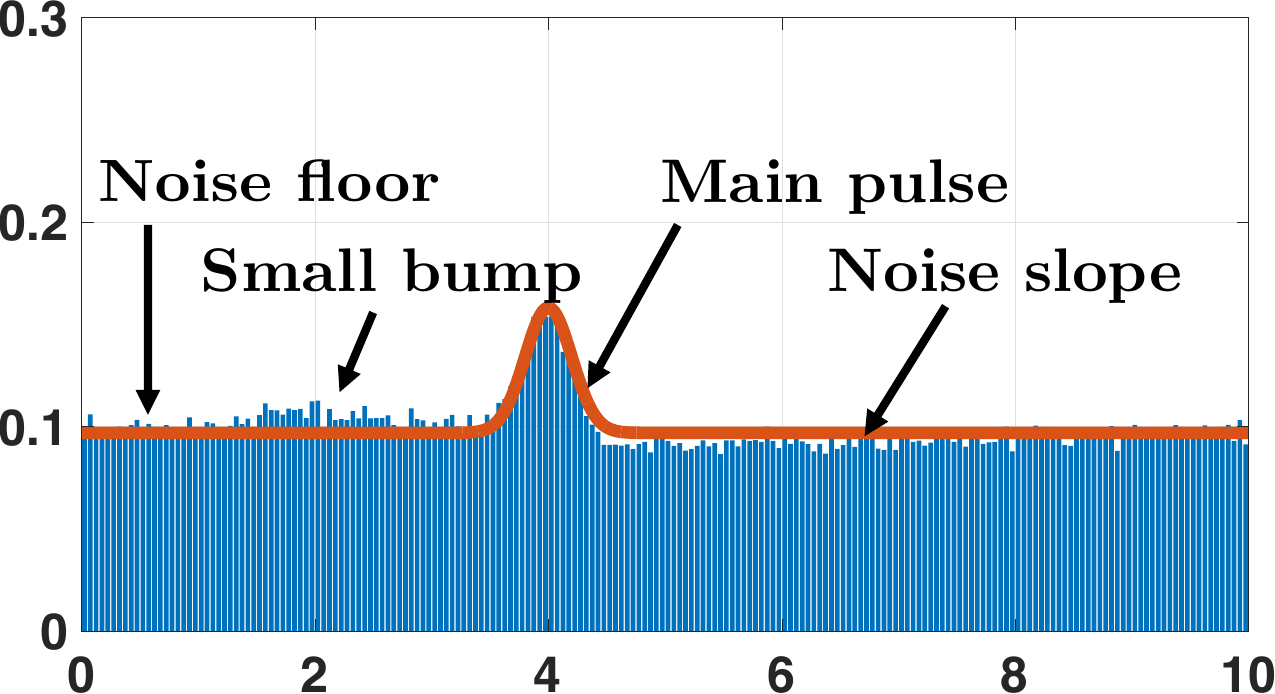}\label{fig: dead time histogram (b)}}
    \caption{Typical dead time distortions. (a) has one main pulse with a small bump ahead of it and a seemingly symmetric vacancy after it. (b) suffers from a high noise floor apart from the main pulse, small bump, and noise slope.}
    \label{fig: dead time histograms}
\end{figure}

By repeating the experiments for $\texttt{iter} = 20$ times and averaging over the histograms, we can empirically obtain the photon registration PDF and visualize the distortions. It is pointed out in \cite{rappDeadTimeCompensation2019} that both the signal and noise levels and the difference between $t_r$ and $t_d$ will influence the distortion shapes. Similar to theirs, we fix the number of cycles $K = 10000$, the dead time $t_d = 7.5$, the depth $\tau = 4$, the half signal pulse width $\sigma_t = 0.2$ and vary $S, B$ between $3.16$ and $0.1$ and $t_r$ between $8$ and $10$. The unit for all time-related variables is 10 ns. Others are counting numbers without units. Fig.~\ref{fig: dead time histograms} shows two examples of distortions.

% Fig~\ref{fig: dead time histogram (a)} contains a single distorted pulse, Fig~\ref{fig: dead time histogram (b)} has one main pulse with a small bump ahead of it and a seemingly symmetric vacancy after it, Fig~\ref{fig: dead time histogram (c)} suffers from a high noise floor, and Fig~\ref{fig: dead time histogram (d)} shares a similar bump with Fig~\ref{fig: dead time histogram (b)} in existence of high noise.

\section{Method}

In this section, we explain our parametric model and the associated EM algorithm to estimate the parameters. We also demonstrate periodic padding to deal with noise slopes.

\subsection{Proposed Parametric Model}

Inspecting Fig.~\ref{fig: dead time histograms}, we notice there is a skewed bell-shaped main pulse in both scenarios. We aim to utilize a Gaussian mixture model (GMM) to represent the distortion with a few Gaussians. However, the flat noise floor in Fig.~\ref{fig: dead time histogram (b)} is tough to be described by Gaussians only. Consequently, a Gaussian-uniform mixture model (GUMM) is introduced to deal with both the distortion and the noise floor.

Assume the photon registration timestamp is a random variable $Y$ with PDF
\begin{equation}
    \label{eq: gaussian uniform}
    f_{Y}(y) = \sum_{m=1}^M \pi_m \, \calN(y; \mu_m, \sigma_m) + \pi_0 \, U(y; 0, t_r),
\end{equation}
where $\calN(y; \mu_m, \sigma_m)$ is the $m$th Gaussian with mean $\mu_m$, variance $\sigma_m^2$, and an associated probability $\pi_m$ and $U(y; 0, t_r)$ is a uniform distribution over $[0, t_r)$ with an associated probability $\pi_0$. $M$ is the total number of Gaussians. In this model, there are $3M+1$ parameters $\vtheta = \bigl\{\pi_0 \cup (\pi_m, \mu_m, \sigma_m)_{m=1}^M \bigl\}$ out of which only $3M$ have to be estimated due to the equality constraint $\sum_{l=0}^M \pi_l= 1$.

Note that the GMM is a special case of the GUMM with $\pi_0 = 0$. In the case of low background noise, we can simply drop the $\pi_0 \, U(y; 0, t_r)$ term and reduce the GUMM to GMM.

\subsection{EM Algorithm}
We consider and derive the EM algorithm for the GUMM in \eqref{eq: gaussian uniform} and that for the GMM can be obtained by the same procedures.

Let $Y$ denote the photon registration timestamp and $X$ denote the corresponding label. Assume we observe $N$ timestamps $\vy = [y_1, y_2, \ldots, y_N]^T$ and the associated label vector $\vx = [x_1, x_2, \ldots, x_N]^T$ ($x_n \in \{0, 1, \ldots, M\}, n = 1, 2, \ldots, N$) indicating which implicit density a timestamp is from, i.e.
\begin{align*}
    \{Y | X = x_n\} \sim
    \begin{cases}
        U(y; 0, t_r) & x_n = 0, \\
        \calN(y; \mu_{x_n}, \sigma_{x_n}) & x_n = 1, 2, \ldots, M.
    \end{cases}
\end{align*}
The task is to estimate $\vtheta = [\pi_0, \pi_1, \mu_1, \sigma_1, \ldots, \pi_M, \mu_M, \sigma_M]^T$ from the observations $(\vy, \vx)$.

If we have full knowledge of $\vx$, it can be easily done by the ML estimation~\cite{10.5555/151045}
\begin{equation*}
    \vthetahat = \argmax{\vtheta \in \Omega} \ \log p(\vy,\vx|\vtheta).
\end{equation*}
Unfortunately, since we do not have $\vx$, the ML estimation from only $\vy$ becomes
\begin{equation*}
    \vthetahat = \argmax{\vtheta \in \Omega} \ \log p(\vy|\vtheta) = \argmax{\vtheta \in \Omega} \ \log \int_\vx p(\vy, \vx|\vtheta) \ d\vx,
\end{equation*}
which is computationally intractable due to the integral over $\vx$. In this case, we can iteratively estimate the parameters utilizing the EM algorithm.

Each update of the EM algorithm is comprised of an E-step and an M-step as follows.
\begin{align}
    \label{eq: Q func}
    & \texttt{E-step:} Q(\vtheta, {\vtheta}^{(k)}) = \E\left[\log p(\vy, \mX | \vtheta) | \mY = \vy, {\vtheta}^{(k)} \right] \\
    & \texttt{M-step:} {\vtheta}^{(k+1)} = \argmax{\vtheta \in \Omega} \ Q(\vtheta, {\vtheta}^{(k)}), \nonumber
\end{align}
where $\vtheta$ is a changing variable, ${\vtheta}^{(k)}$ is the current estimation, and ${\vtheta}^{(k+1)}$ is the next estimation. Instead of directly maximizing the objective function $\log p(\vy|\vtheta)$, we construct and maximize the surrogate function $Q(\vtheta, {\vtheta}^{(k)})$. The theoretical basis for the convergence of the EM algorithm can be found in~\cite{baumMaximizationTechniqueOccurring1970,dempsterMaximumLikelihoodIncomplete1977}.

To implement the EM for the GUMM, instead of tediously solving \eqref{eq: Q func}, a shortcut exists from Equation (12.34) in \cite{doi:10.1137/1.9781611977134}. Let $T(\vy, \vx)$ be a sufficient statistic of $p(\vy, \vx|\vtheta)$, then the ML estimate of $\vtheta$ is given by $\vthetahat_{ML} = f(T(\vy, \vx))$ for some function $f(\cdot)$. Since $p(\vy, \vx|\vtheta)$ belongs to the exponential family, it can be shown that one update of the EM algorithm is $\vtheta^{(k+1)} = f(\widebar{T}(\vy))$, where $\widebar{T}(\vy) = \E[T(\vy, X|Y = \vy, \vtheta^{(k)})]$~\cite{doi:10.1137/1.9781611977134}.

Accordingly, the E-step includes
\begin{align}
\label{eq: E_step}
    \widebar{N}_l & = \sum_{n=1}^{N} \Pr(X_n = l | Y = y_n, \vtheta^{(k)}) \\
    \overline{t}_{1, m} & = \sum_{n=1}^{N} y_n \Pr(X_n = m | Y = y_n, \vtheta^{(k)}) \\
    \overline{t}_{2, m} & = \sum_{n=1}^{N} y_n^2 \Pr(X_n = m | Y = y_n, \vtheta^{(k)}),
\end{align}
where $\widebar{N}_l, \overline{t}_{1, m}, \overline{t}_{2, m}$ are expected values of the sufficient statistics of $\pi_l, \mu_m, \sigma_m^2$ respectively, $l = 0, 1, \ldots, M$, and $m = 1, 2, \ldots, M$. The M-step includes
\begin{equation}
\label{eq: M_step}
    \pihat_l = \frac{\widebar{N}_l}{N} ,\ \muhat_m = \frac{\overline{t}_{1, m}}{\widebar{N}_m} ,\ \sigmahat_m^2 = \frac{\overline{t}_{2, m}}{\widebar{N}_m} - \frac{\overline{t}_{1, m}^2}{\widebar{N}_m^2}.
\end{equation}
% \begin{align*}
%     \pihat_m & = \frac{\widebar{N}_m}{N} \\
%     \muhat_m & = \frac{\widebar{t}_{1, m}}{\widebar{N}_m} \\
%     \sigmahat_m^2 & = \frac{\widebar{t}_{2, m}}{\widebar{N}_m} - \frac{\widebar{t}_{1, m}^2}{\widebar{N}_m^2}.
% \end{align*}
What remains is to calculate the posterior probability
\begin{align}
    \label{eq: posterior}
    & \Pr(X_n = l | Y = y_n, \vtheta^{(k)}) = \frac{\Pr(X_n = l , Y = y_n| \vtheta^{(k)})}{\Pr(Y = y_n| \vtheta^{(k)})} \notag \\
    & =
    \begin{dcases}
        \frac{\pi_0^{(k)} \, (1/t_r)}{\sum_{j = 0}^M \Pr(X_n = j , Y = y_n| \vtheta^{(k)})} & \text{if }l = 0, \\
        \frac{\pi_l^{(k)} \, N(y_n; \mu_l^{(k)}, \sigma_l^{(k)})}{\sum_{j = 0}^M \Pr(X_n = j , Y = y_n| \vtheta^{(k)})} & \text{otherwise},
    \end{dcases}
\end{align}
where the denonminator $\sum_{j = 0}^M \Pr(X_n = j , Y = y_n| \vtheta^{(k)})$ is
\begin{equation*}
    \pi_0^{(k)} \, (1/t_r) + \sum_{m = 1}^M \pi_m^{(k)} \, N(y_n; \mu_m^{(k)}, \sigma_m^{(k)}).
\end{equation*}

The parameters can be recursively estimated by exploiting from \eqref{eq: E_step} to \eqref{eq: posterior}. Upon convergence, we get an approximated photon registration PDF parameterized by $\vthetahat$.

\subsection{Periodic Padding}
In some situations where the asymmetric noise slopes are difficult to model using either GMM or a uniform plateau, we introduce a periodic padding trick.

\begin{figure}[tbp]
     \centering
         \includegraphics[width=0.48\textwidth]{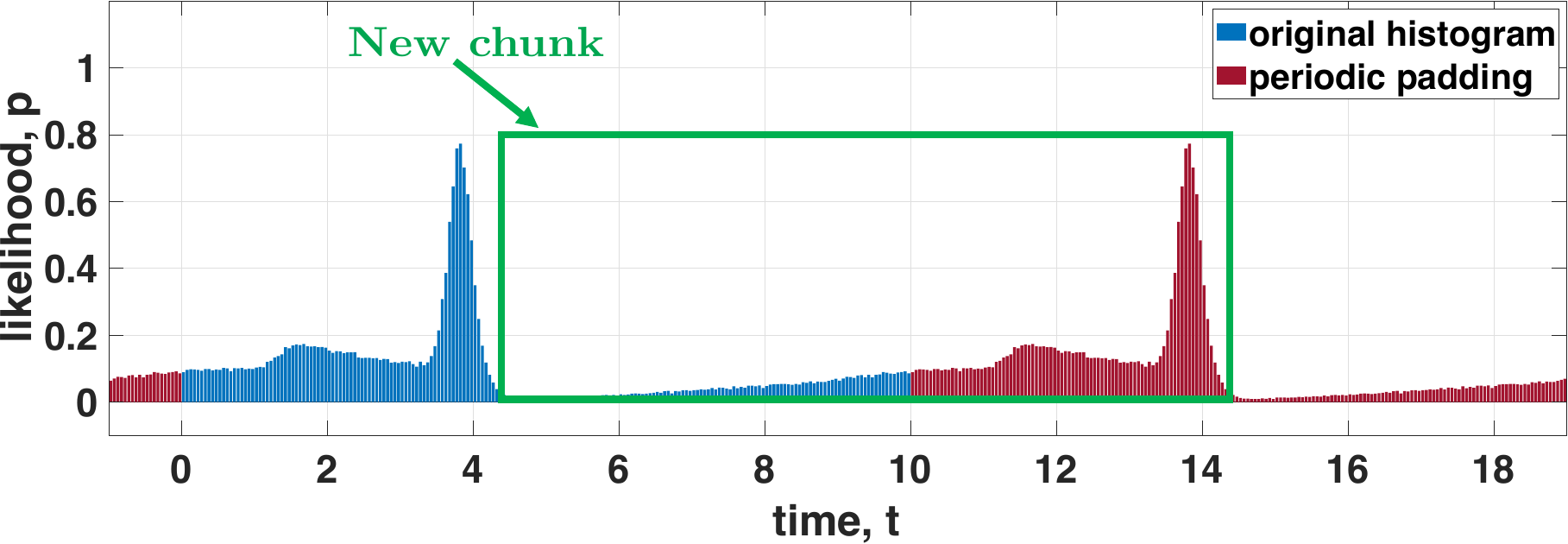}
    \caption{Illustration of periodic padding. We periodically pad the original blue histogram with the red ones. A new chunk is selected as the EM estimation input for boundary consistency.}
    \label{fig: padding}
\end{figure}

The observation is that the empirical histogram can be extended and replicated along its repetition period. We can periodically pad the histogram and re-select a chunk with the length $t_r$ to take advantage of the consistency between the left and right boundary. The idea is illustrated in Fig.~\ref{fig: padding}.

The periodic padding is complementary to our pipeline because we can shift the timestamp values to the new chunk, run EM to estimate the parameters, and then transform the shape back to the original domain.

\section{Experimental Results}\label{Sec: Exp}

\subsection{Data and Metric}
We consider $4$ imaging scenarios as outlined from~Fig.\labelcref{fig: single pulse,fig: noise,fig: sig of padding}. For each case, we simulate $20$ histogram realizations and average them as the approximated ground truth PDF. The histogram binning resolution is $0.5$ ns. Meanwhile, all timestamps among the $20$ iterations are stored as the measurement data $\vy$ for the EM. The typical number of timestamps used is around $9000$. We run the EM for $\texttt{EM-iter} = 50$ times to ensure the convergence. If the model complexity grows, e.g. more Gaussians are used, we increase $\texttt{EM-iter} = 80$ times.

After the EM, we obtain a mixture PDF with the estimated parameters $\vthetahat$. We sample values at the corresponding positions of the histogram bins from the PDF and calculate the mean squared error ($\MSE$) between the predicted PDF and the averaged histogram to evaluate the accuracy of our method.

\subsection{Fitting Performance}
\begin{figure}[tbp]
     \centering
         \includegraphics[width=0.24\textwidth]{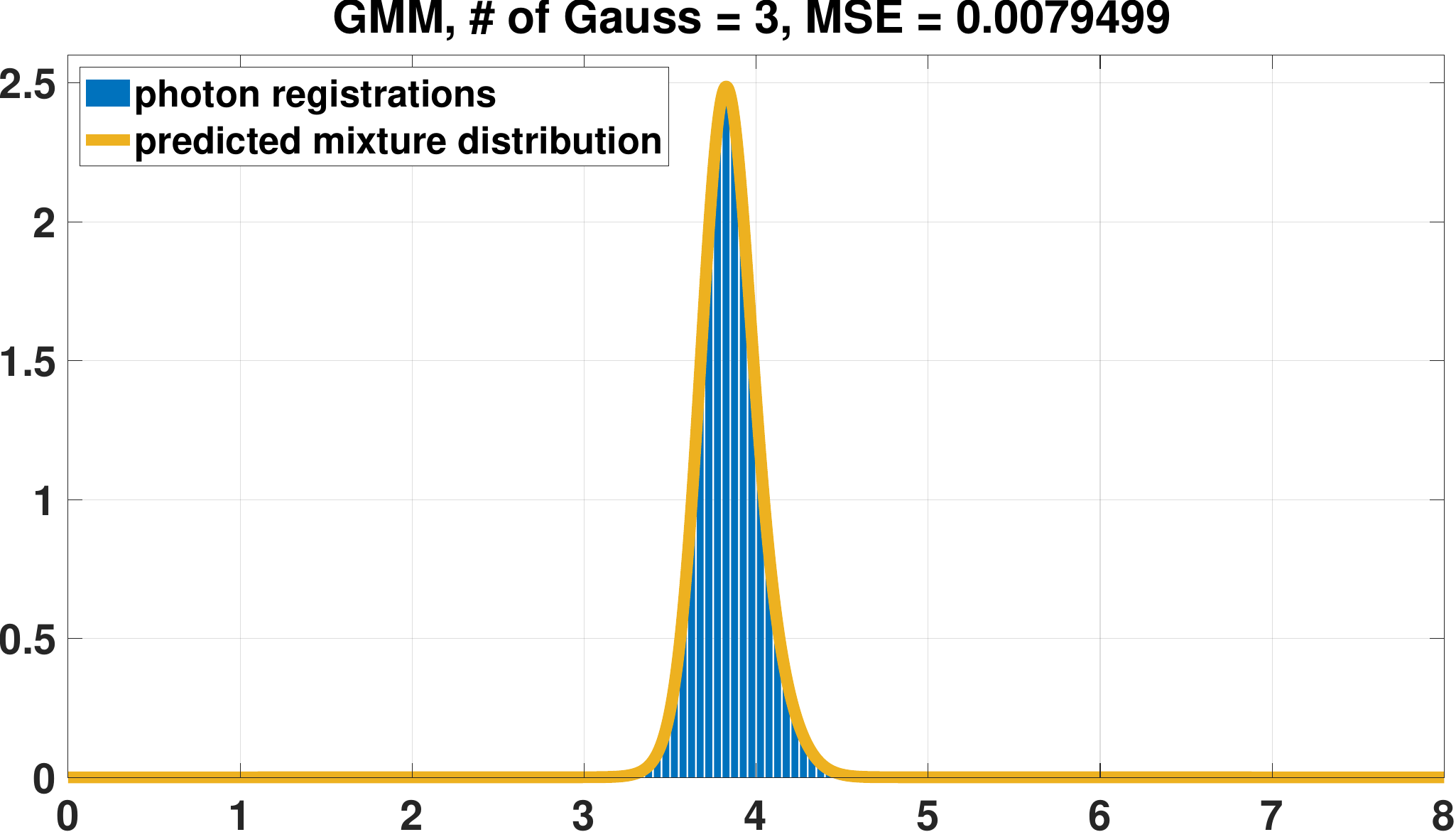}
    \caption{Fitting of a single distorted pulse.}
    \label{fig: single pulse}
\end{figure}

For a single distorted pulse in Fig.~\ref{fig: single pulse}, our GMM model matches the histogram with $3$ Gaussians in $50$ EM iterations.

\subsubsection{Significance of GUMM}
\begin{figure}[tbp]
     \centering
     \subfloat[]{\includegraphics[width=0.24\textwidth]{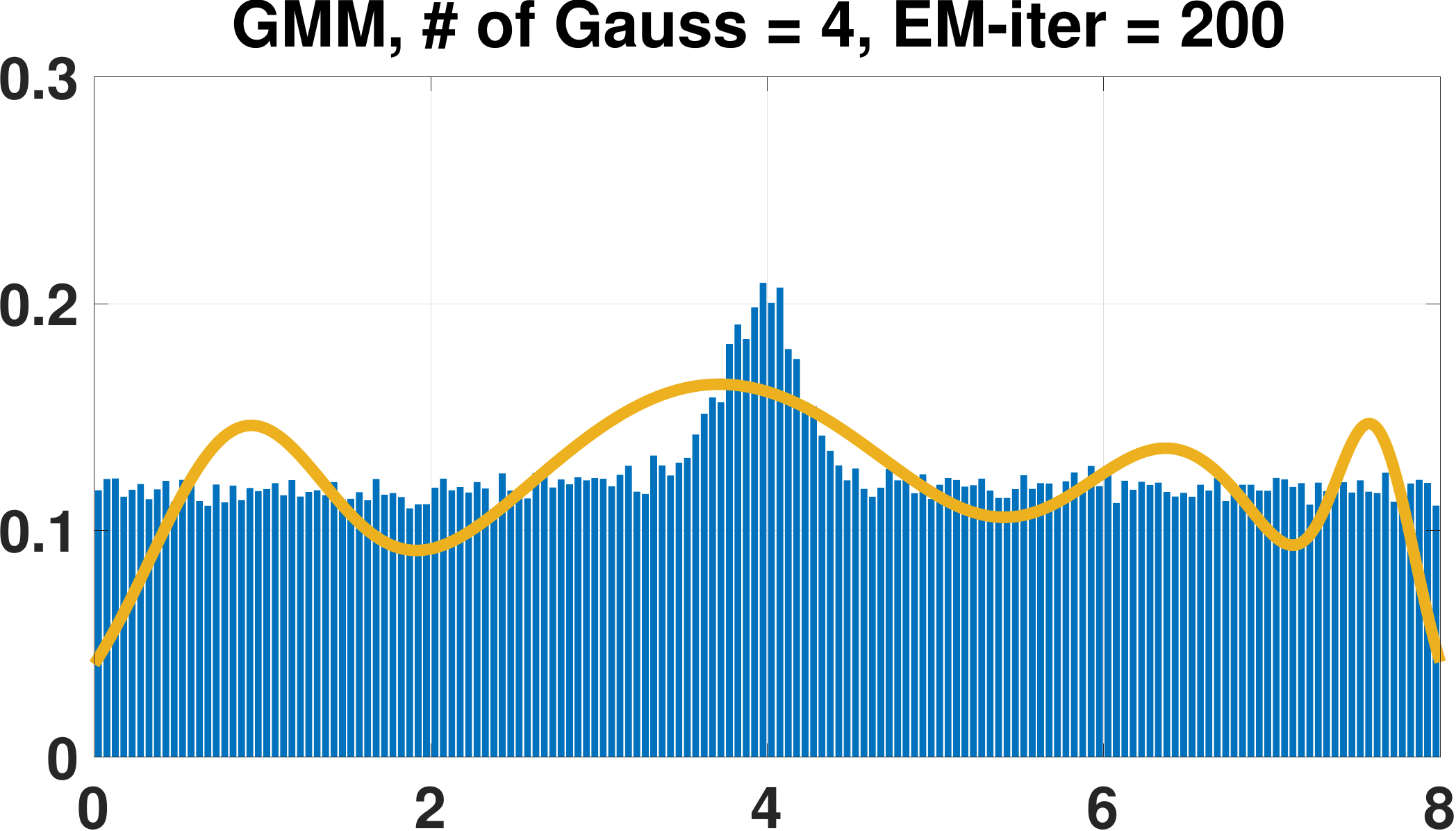}\label{fig: noise GMM}}
     \hfill
     \subfloat[]{\includegraphics[width=0.24\textwidth]{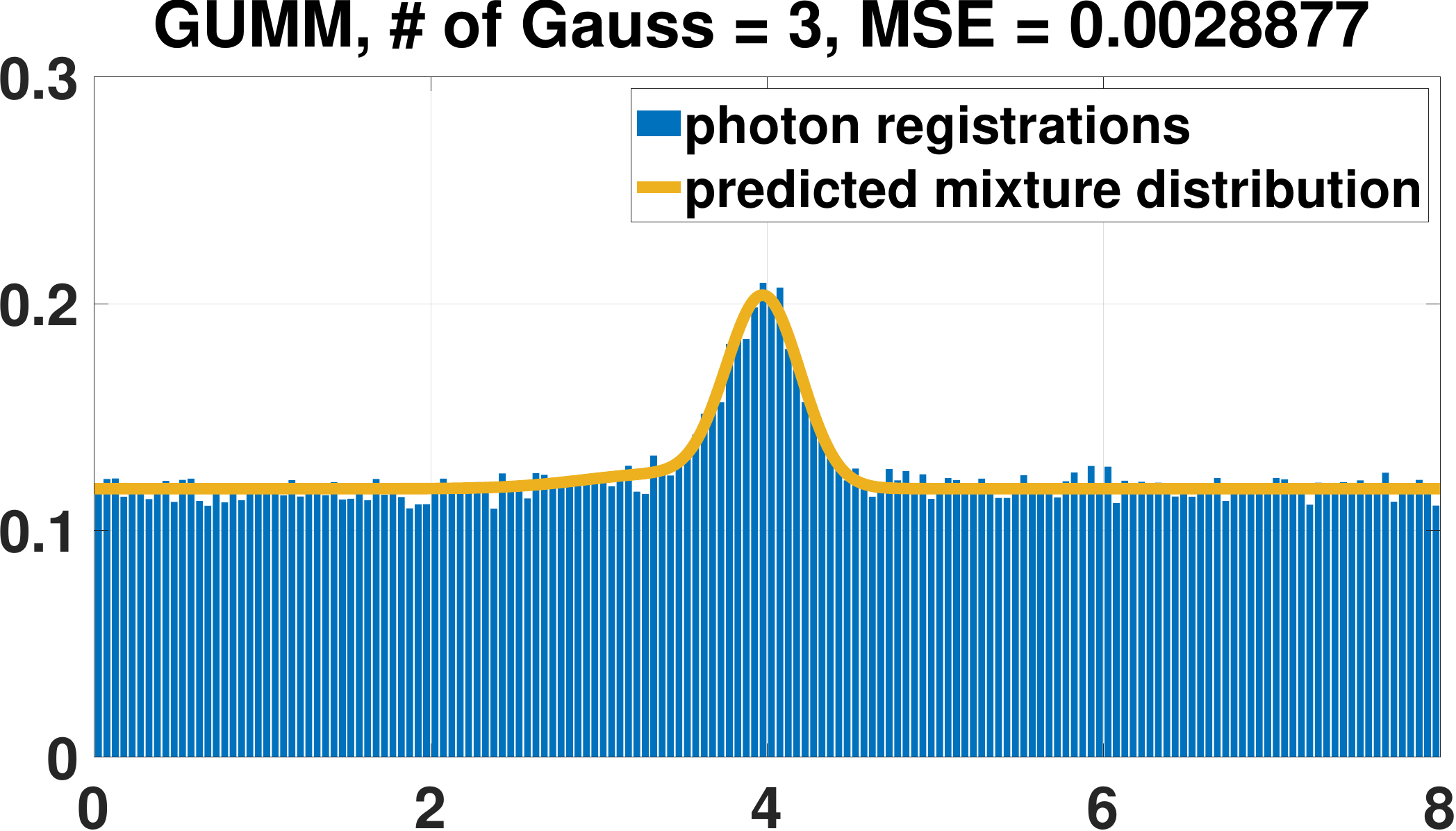}\label{fig: noise GUMM}}
    \caption{Fitting of a single pulse with a high noise floor. (a) Failure of GMM. (b) Success of GUMM.}
    \label{fig: noise}
\end{figure}

In Fig.~\ref{fig: noise}, we verify the necessity of GUMM for high noise levels. The GMM is incapable of modeling the high noise floor, as in Fig.~\ref{fig: noise GMM}. But with an extension to the GUMM, Fig.~\ref{fig: noise GUMM} achieves a good match with fewer Gaussians ($3$) and iterations ($50$).

\subsubsection{Significance of Periodic Padding}
\begin{figure}[tbp]
     \centering
     \subfloat[]{\includegraphics[width=0.24\textwidth]{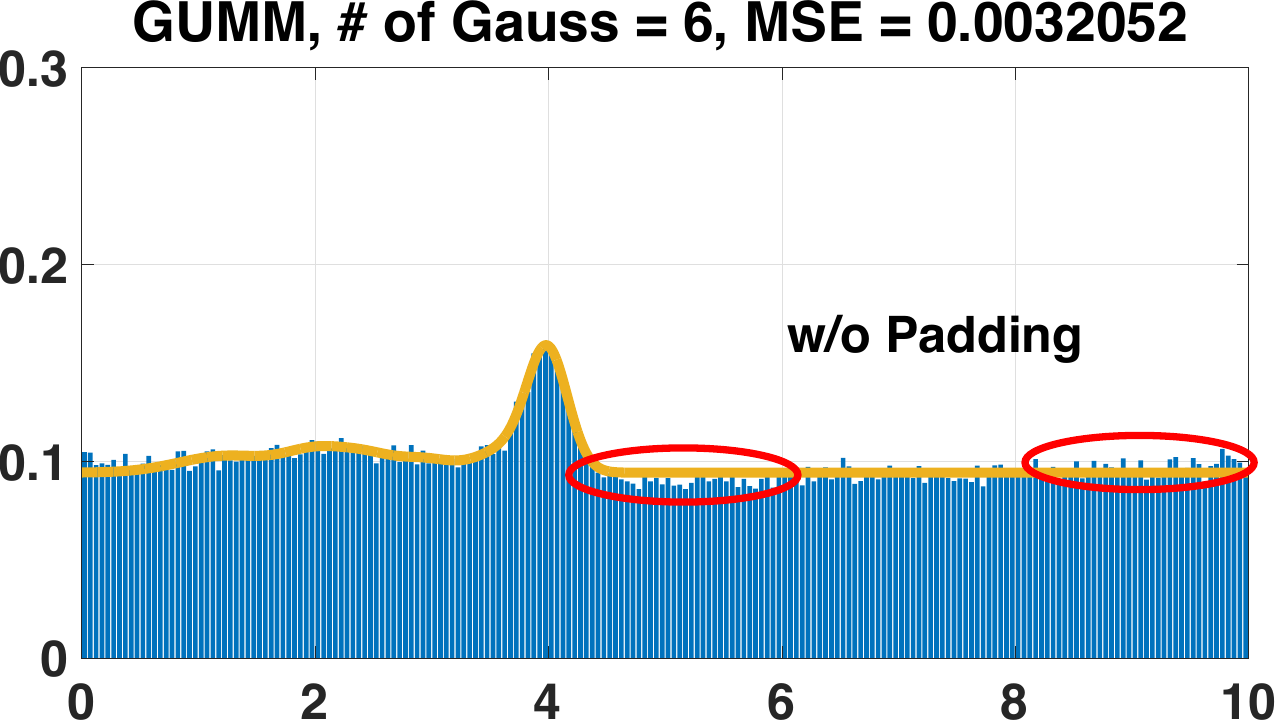}\label{fig: bump noise w/o padding}}
     \hfill
     \subfloat[]{\includegraphics[width=0.24\textwidth]{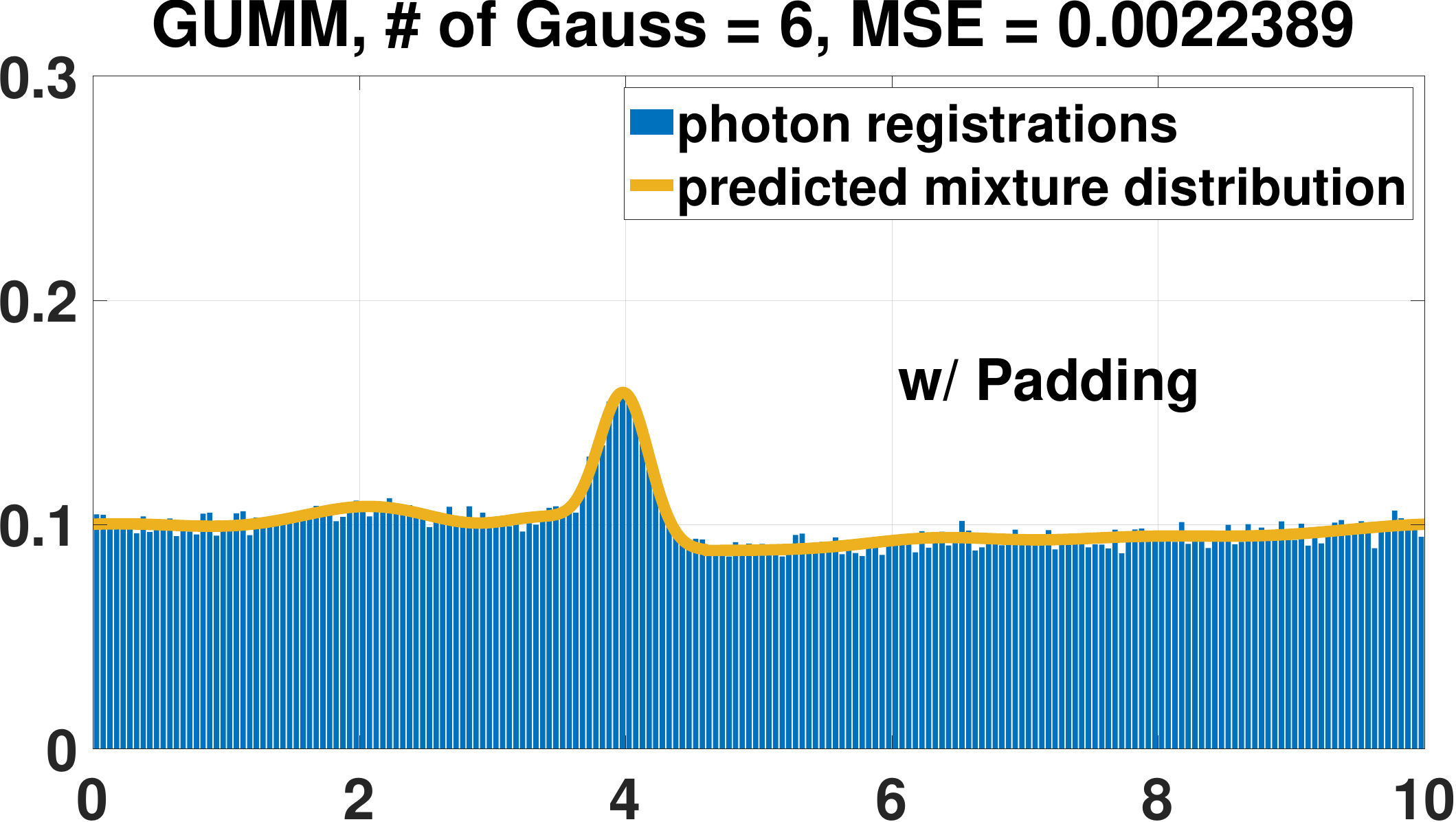}\label{fig: bump noise w/ padding}}

     \subfloat[]{\includegraphics[width=0.24\textwidth]{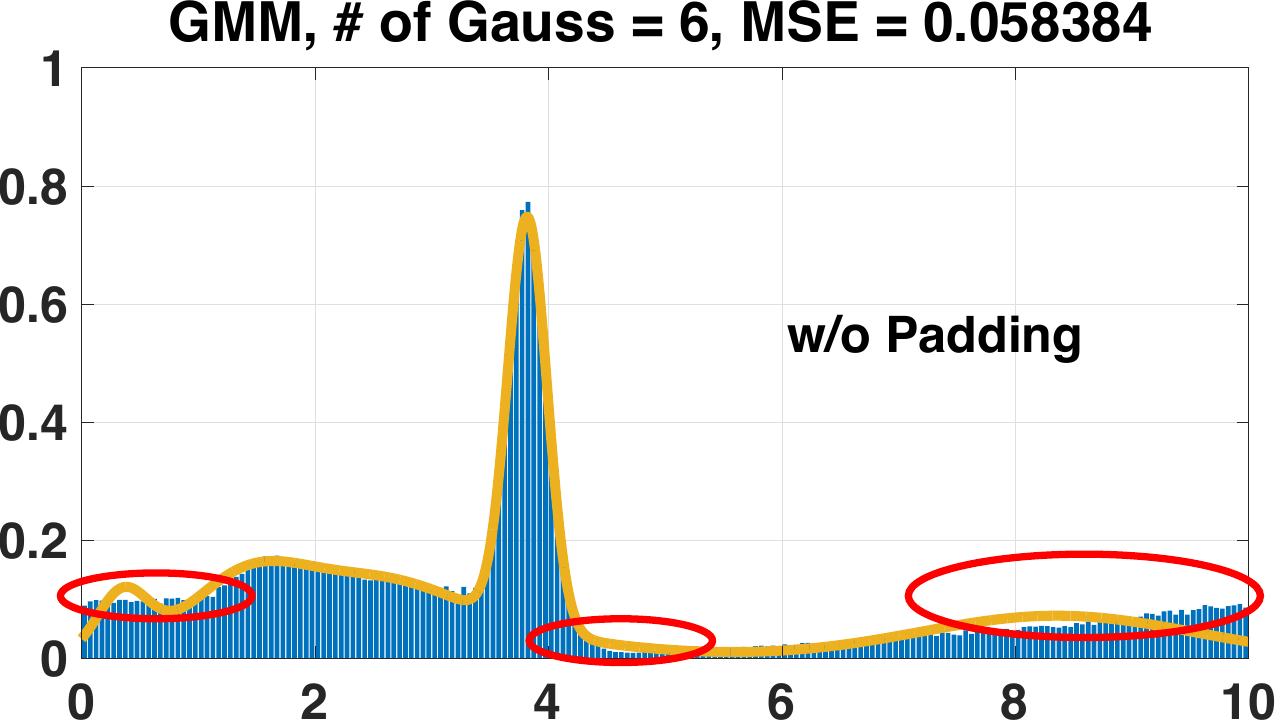}\label{fig: bump w/o padding}}
     \hfill
     \subfloat[]{\includegraphics[width=0.24\textwidth]{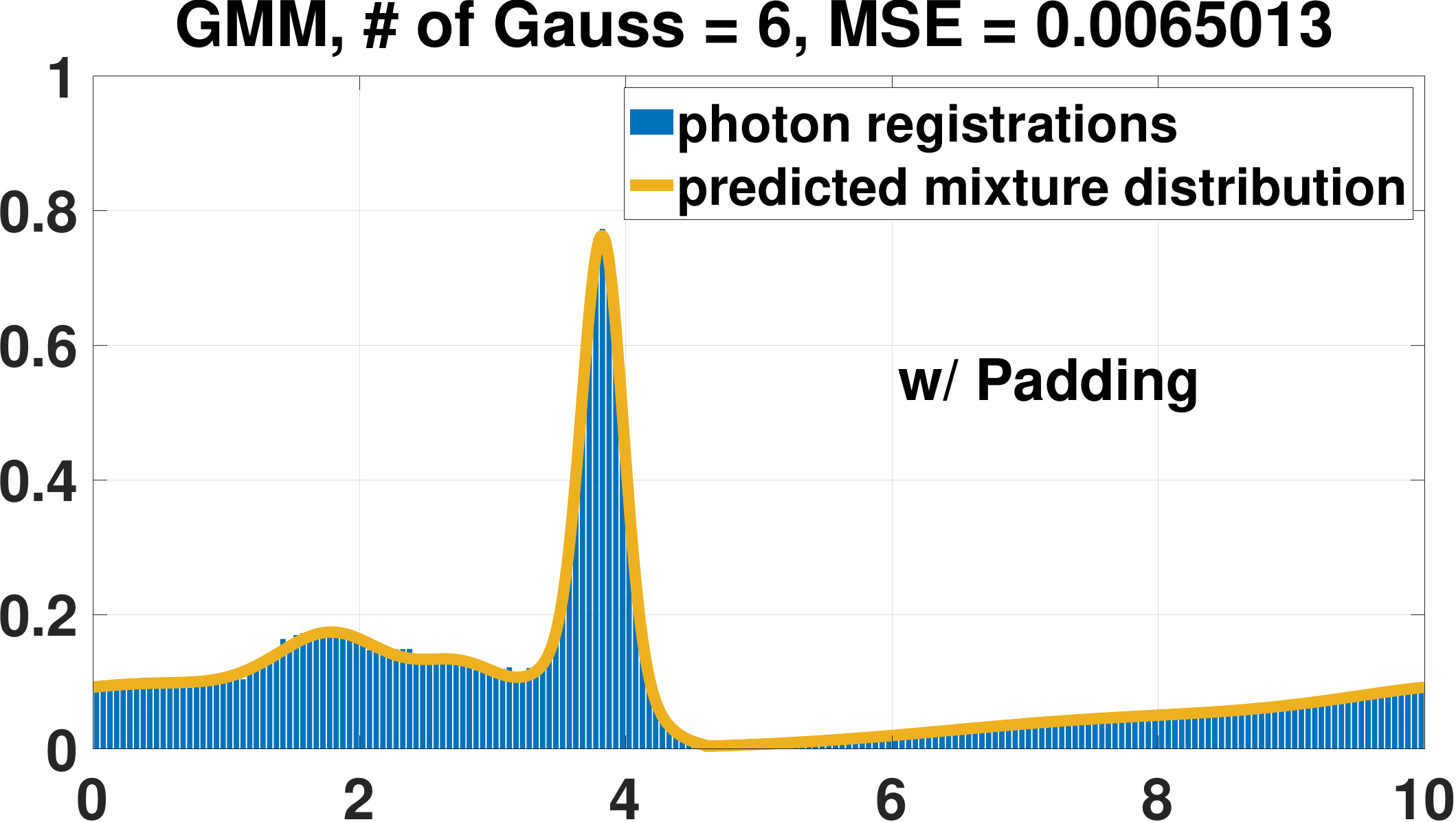}\label{fig: bump w/ padding}}
    \caption{Fitting of a small bump ahead of the main peak and a noise slope after it both with and without high noise floors. (a) Failure of GUMM w/o padding. (b) Success of GUMM w/ padding. (c) Failure of GMM w/o padding. (d) Success of GMM w/ padding.}
    \label{fig: sig of padding}
\end{figure}

Periodic padding plays a vital role in fitting the small bump and the noise slope as in Fig.~\ref{fig: sig of padding}. Without the padding, though Fig.~\ref{fig: bump noise w/o padding} has a decent capture of the two peaks, it fails to describe the vacancy in the valley and the slope on the right boundary, which are fitted well in Fig.~\ref{fig: bump noise w/ padding}. Similarly, the GMM cannot predict the valley and the boundaries well without padding as shown in Fig.~\ref{fig: bump w/o padding}. With padding, however, the match is almost perfect in Fig.~\ref{fig: bump w/ padding}.

\subsubsection{Quantitative results}
The $\MSE$s of our predictions in the four cases are summarized in Table~\labelcref{table:1,table:2}. The bold numbers correspond to successful results from~Fig.\labelcref{fig: single pulse,fig: noise,fig: sig of padding}.

\begin{table}[tbp]
\caption{$\MSE$ of single pulse (Fig.~\ref{fig: single pulse}) and high noise (Fig.~\ref{fig: noise})}
\begin{center}
\begin{tabular}{cccc}
 & & \multicolumn{2}{c}{\textbf{$\#$ of Gaussians used}} \\
\toprule
\textbf{Scenario} & \textbf{Model} & \textbf{2} & \textbf{3} \\
\midrule
Single pulse (Fig.~\ref{fig: single pulse}) & GMM & 0.18994 & \textbf{0.00795} \\
High noise (Fig.~\ref{fig: noise}) & GUMM & 0.00291 & \textbf{0.00289} \\
\bottomrule
\end{tabular}
\label{table:1}
\end{center}
\end{table}

\begin{table}[tbp]
\caption{$\MSE$ of bump ($2$nd row Fig.~\ref{fig: sig of padding}) and bump w/ noise ($1$st row Fig.~\ref{fig: sig of padding})}
\begin{center}
\begin{tabular}{>{\centering\arraybackslash}m{1.9cm}>{\centering\arraybackslash}m{0.7cm}>{\centering\arraybackslash}m{0.9cm}>{\centering\arraybackslash}m{0.9cm}>{\centering\arraybackslash}m{0.9cm}>{\centering\arraybackslash}m{0.9cm}}
 & & & \multicolumn{3}{c}{\textbf{$\#$ of Gaussians used}} \\
\toprule
\textbf{Scenario} & \textbf{Model} & \textbf{Padding} & \textbf{4} & \textbf{5} & \textbf{6} \\
\midrule
Bump & \multirow{2}{*}{GMM} & \xmark & 0.08295 & 0.06609 & 0.05838 \\
($2$nd row Fig.~\ref{fig: sig of padding}) & & \cmark & 0.02130 & 0.01294 & \textbf{0.00650} \\
Bump w/ noise & \multirow{2}{*}{GUMM} & \xmark & 0.00328 & 0.00321 & 0.00320 \\
($1$st row Fig.~\ref{fig: sig of padding}) & & \cmark & 0.00241 & 0.00228 & \textbf{0.00224} \\
\bottomrule
\end{tabular}
\label{table:2}
\end{center}
\end{table}

To model more complicated distortion shapes, more Gaussians are required for accurate fitting. As shown in Table~\ref{table:2}, we need at most $6$ Gaussians ($18$ parameters) to model all cases, which is much simpler than the Markov chain approximation. Table~\ref{table:2} also demonstrates that the prediction of the bump cases is more precise and efficient with periodic padding.

\section{Conclusion}
In this paper, we present a parametric model to estimate the stationary photon registration PDF affected by the dead time distortions in modern LiDAR setups. We include a uniform density in the mixture model to deal with extremely noisy cases and propose a periodic padding technique to overcome the small bump and the noise slope. The parameters are computed iteratively using the EM algorithm. The experimental results show that our proposed model, the associated EM, and padding can efficiently and accurately predict the photon registration PDF with at most eighteen parameters and eighty iterations. Compared to the state-of-the-art Markov chain model, our method is continuous, differentiable, and more concise, which is ideal for deep learning and fast timestamp sampling applications. It will further speed up the acquisition rate in modern TCSPC systems, including but not limited to SP-LiDAR.

\bibliographystyle{IEEEtran}
\bibliography{IEEEabrv,main}

\end{document}